\documentclass[conference]{IEEEtran}
\usepackage{multirow}
\usepackage{amssymb}
\usepackage{ifpdf}
\usepackage{cite}
\usepackage[pdftex]{graphicx}
\usepackage[cmex10]{amsmath}
\usepackage{algorithmic}
\usepackage{array}
\usepackage{color,soul}
\usepackage{mdwmath}
\usepackage{mdwtab}
\usepackage{eqparbox}
\usepackage{caption}
\usepackage{subcaption}
\usepackage{fixltx2e}
\usepackage{stfloats}
\usepackage{url}
\usepackage{mdframed}
\usepackage{hyperref}

\newcommand{\etal}[0]{et~al{.}}
\newenvironment{finding}{\begin{mdframed}\noindent~}{\end{mdframed}}

\title{Threat Models over Space and Time:\\ A Case Study of E2EE Messaging Applications}
\author{
\IEEEauthorblockN{Partha Das Chowdhury\IEEEauthorrefmark{1}, Maria Sameen\IEEEauthorrefmark{1}, Jenny Blessing\IEEEauthorrefmark{2}, Nicholas Boucher\IEEEauthorrefmark{2}, Joseph Gardiner\IEEEauthorrefmark{1}, Tom Burrows\IEEEauthorrefmark{2}, \\Ross Anderson\IEEEauthorrefmark{2}\IEEEauthorrefmark{3}, Awais Rashid\IEEEauthorrefmark{1}}
\IEEEauthorblockA{\IEEEauthorrefmark{1}University of Bristol, UK \emph{\{partha.daschowdhury, maria.sameen, joe.gardiner, awais.rashid\}@bristol.ac.uk}}
\IEEEauthorblockA{\IEEEauthorrefmark{2}University of Cambridge, UK. \emph{\{jenny.blessing, nicholas.boucher, ross.anderson\}@cl.cam.ac.uk, tom@tpmb.uk}}
\IEEEauthorblockA{\IEEEauthorrefmark{3}University of Edinburgh, UK \emph{ross.j.anderson@ed.ac.uk}}
}
\begin{document}

\IEEEoverridecommandlockouts

\maketitle

\newpage
\begin{abstract}
  Threat modelling is one of the foundations of secure systems engineering and must take heed of the context within which systems operate. In this work, we explore the extent to which real-world systems engineering reflects a changing threat context. We examine the desktop clients of six widely used end-to-end-encrypted mobile messaging applications to understand the extent to which they adjusted their threat model over space (when enabling clients on new platforms, such as desktop clients) and time (as new threats emerged). We experimented with short-lived adversarial access against these desktop clients and analyzed the results using two popular threat elicitation frameworks, STRIDE and LINDDUN. The results demonstrate that system designers need to track threats in the evolving context within which systems operate and, more importantly, mitigate them by rescoping trust boundaries so that they remain consistent with administrative boundaries. A nuanced understanding of the relationship between trust and administration is vital for robust security, including the provision of safe defaults.

  % among extensions of existing systems
  
\end{abstract}

\section{Introduction}\label{sec:intro}

Threat modelling has become an integral part of secure software development. Using a framework such as STRIDE (Spoofing, Tampering, Repudiation, Information disclosure, Denial of service, Elevation of privilege) is a key step within the Microsoft Security Development Lifecycle (SDL)~\cite{howard2006security} to analyze application information flows against key classes of threats. Threat modelling is also a recommended best practice by OWASP~\cite{owasp_guide} and within Agile~\cite{safecode} and DevOps processes~\cite{secdevops}. Researchers have developed similar frameworks to systematically analyze threats to user privacy when developing software applications~\cite{deng2011privacy}.

However, threat modelling cannot be a one-off activity. Any entity doing so as a one-off activity is not following documented SDL best practices as threats evolve and new attacks come to light, developers must continuously reassess applications against them. Furthermore, adding new features to an application creates new information flows, which in turn can cause trust boundaries to shift, as can also happen as systems acquire additional hardware or software components or third-party services. The recommended best practice is to do threat modelling ``little and often''~\cite{gumbley2020}.

In this paper, we analyze whether the threat models that underpin security and privacy aspects of applications evolve through space (as new features are added) and time (as the understanding of threats changes). Our case study consists of end-to-end-encrypted (E2EE) mobile messaging applications (such as Signal~\footnote{We refer to the Signal messenger application here, rather than the Signal protocol, and will continue to distinguish between the two throughout the paper.} and WhatsApp), which have found widespread adoption amongst users aiming to protect the privacy of their communications and mitigate large-scale surveillance~\cite{signal_user,e2ee_user}. E2EE mobile apps were conceived to protect message contents from potential eavesdroppers on the communication channel (which includes the messaging provider itself). Most of these platforms have since also launched desktop clients to make it easier for users to handle larger, more complex messages, and communicate across multiple devices.

In the meantime, security and privacy threats against which users need to be protected have also evolved. For instance, the mobile app messaging threat model was largely predicated on an eavesdropper on the communication channel. However, research into intimate partner violence has highlighted that abusers often use monitoring technology or shared devices to surveil and exert control over victims~\cite{chatterjee2018, tseng2020}. In this scenario, the threat actor is not remote but has direct physical access to the victim's devices. For an abuse victim, the persistence of access, which may last long after they have left their abuser, is of serious concern. 

There are other contexts in which even short-lived access to a desktop client can pose potential threats:

\begin{itemize}
\item \textbf{Official searches:} Border and customs officials sometimes search travellers' devices and make copies of data stored on them. Potential exploitation of the access to E2EE messaging enabled by desktop clients would pose a serious threat to activists and journalists -- a major concern in the context of the Russian invasion of Ukraine and the use of E2EE messaging applications to coordinate activities by defenders and for protecting free speech. The leakage of medical communications is a further concern following the overturning of Roe v. Wade in the US.
  
\item \textbf{Shared devices:} Many households share devices, especially desktop computers and laptops. There is a high likelihood that individuals living in the same household could gain at least temporary access to each other's machine or device backups. The security and privacy provided by E2EE messaging should avoid leakage of sensitive information in domestic contexts where there may be strife or abuse. 

\item \textbf{Corporate Machines:} Machines provided by an employer are often managed remotely, with company sysadmins having full access. Any desktop clients of E2EE messaging applications must ensure that the communications are protected from adversarial sysadmins.
\end{itemize}

\begin{figure}
  \centering
       \fbox{\includegraphics[width=0.95\columnwidth]{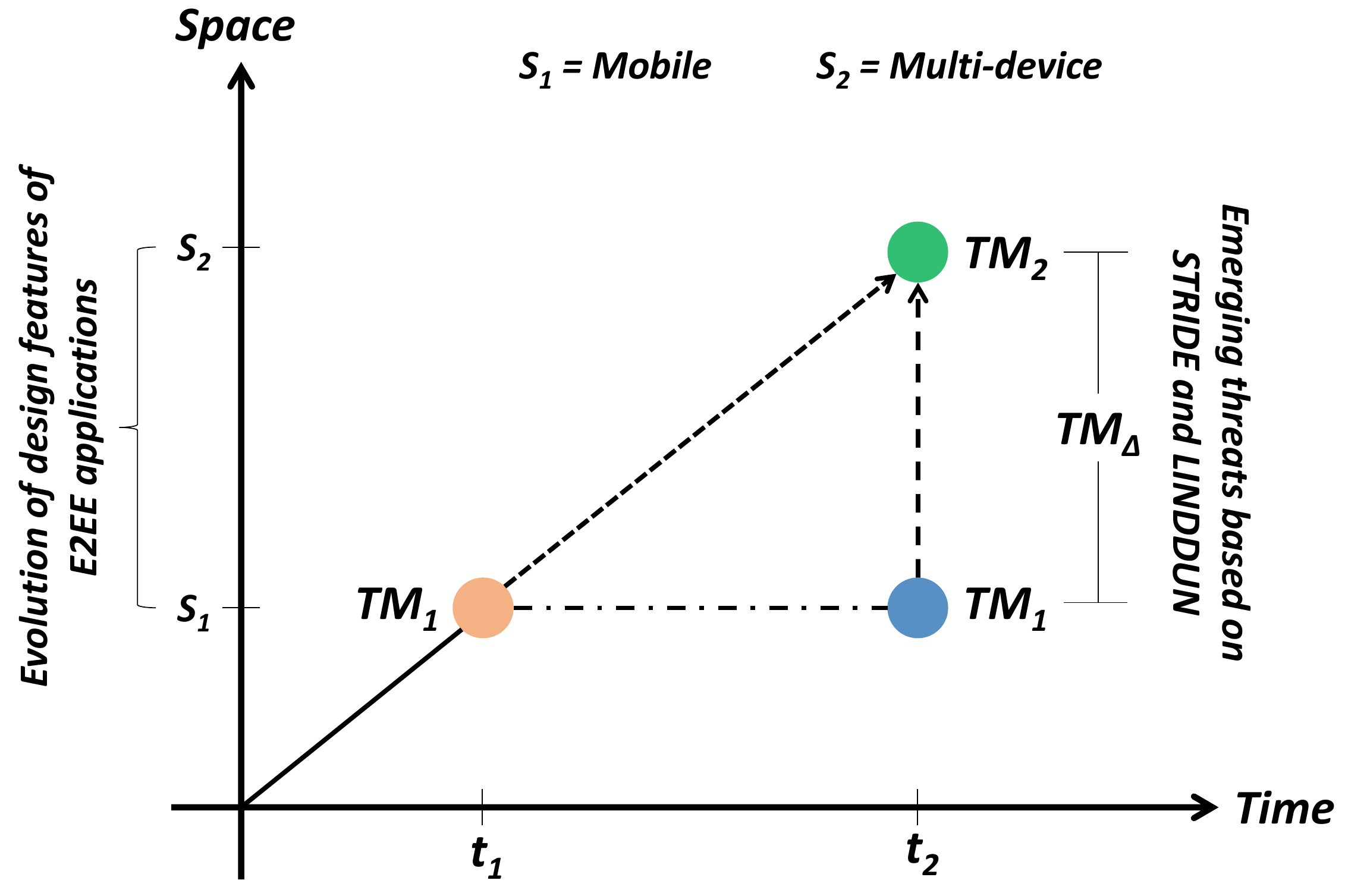}}
   \caption{Evolution of Threat Models}
   \label{fig:intro-threatmodels}
\end{figure}

Figure~\ref{fig:intro-threatmodels} represents a conceptualization of the importance of evolving threat models over space and time. In order to explore if this happens in practice, we systematically analyze six major E2EE messaging applications: Signal, WhatsApp, Element, Viber, Wickr Me, and Telegram. We start from the original threat model ($TM\textsubscript{1}$) of these applications, i.e., a mobile app client $s\textsubscript{1}$ with a remote attacker at time $t\textsubscript{1}$. We then develop a second ($TM\textsubscript{2}$) at the intersection of the expanded feature space i.e. the desktop client in its operating context (shared and/or managed device) $s\textsubscript{2}$ at time $t\textsubscript{2}$. Using an experimental test setup, we then simulate adversarial short-lived access ($TM\textsubscript{2}$) to the desktop clients of each of the six applications. The resultant compromise (if any) is mapped with respect to security (using the STRIDE threat modelling approach~\cite{howard2006security}) and privacy (using the LINDDUN threat modelling approach~\cite{deng2011privacy}) to elicit security \& privacy threats respectively. The resultant security \& privacy threats comprise the net evolution ($TM\textsubscript{$\Delta$}$) for each application. Our analysis shows the applications evolve their threat model to varying degrees to mitigate the threats resulting from such adversarial short-lived access.

We argue based on our investigation that threat models (and hence protection mechanisms informed by them) need to evolve in space and time as threats change. For some desktop clients, $TM\textsubscript{$\Delta$}$ reveals their vulnerability to \emph{spoofing, repudiation, information disclosure and elevation of privilege} in the face of short-lived adversarial access. These vulnerabilities, in turn, give rise to two kinds of privacy leakage: \emph{linkability} of information and \emph{identifiability} of the communicating parties. % Their documentation explains the protection mechanisms they employ against these threats by an eavesdropper, but their security guarantees fail against short-lived adversarial access to their desktop clients. This reflects their persistence with the same assumptions about potential attackers between the mobile applications and their desktop clients.
Our analysis also highlights that recognition of the change in the threat context is useful but not enough in itself unless backed by appropriate countermeasures. %  as we observed for some applications. Mere recognition can perhaps protect against \emph{elevation of privilege} or can restrict an attacker by duration leaving potential victims vulnerable to other threats and places unnecessary burden on the users. $TM\textsubscript{$\Delta$}$ reveals an absence of the threats for applications which acknowledge the change in the threat context (short-lived adversarial access to desktop clients) and back them with appropriate countermeasures. 

%  For instance, Signal and WhatsApp both extend the Signal protocol, yet the latter notifies the user when an additional device is added to their account. This reflects the differing perceptions of an attacker held by the designers and developers of these applications. Once we captured the $TM\textsubscript{$\Delta$}$ for each application, we reached out to the respective organizations to understand the reasons behind their different perceptions of an attacker and protection mechanisms. 

% Following appropriate ethics approval, We reached out to developers of all six applications to seek interviews. 

\section{Background - E2EE Messaging Applications \& Threat Frameworks}
\label{background}
%\begin{itemize}
 %   \item What are E2EE encrypted messengers
 %  \item What are the common ones. What protocols do they use?
 %   \item Multi-device support in E2EE
 %   \begin{itemize}
 %       \item What apps do what
 %       \item How do they work
 %   \end{itemize}
% \end{itemize}

% Security primitives used in E2EE applications reflect the threat models they consider.
In this section, we tease out the root assumptions underpinning the security of the E2EE applications we investigate. We summarize the key security properties of these applications and their desktop clients in Table \ref{tab:summary-e2ee}. This is followed by a description of the security and privacy threat modelling frameworks we use for this investigation --- STRIDE~\cite{howard2006security} and LINDDUN~\cite{deng2011privacy}.

\subsection{E2EE Messaging Applications}
Every installation is tied to a particular user identity. This identity is then used as the root of trust to communicate securely with other participants through the service and to configure additional devices for the same account. The steps taken to generate a long-term identity key and corresponding short-lived asymmetric key pairs are as follows:
\begin{enumerate}
    \item Every communication node generates its own long-term public-private key pair -- the identity key \(IK\). Any principal that can prove possession of the secret component of \(IK\) is considered the legitimate owner of the account connected to this identity key.
    \item Ephemeral asymmetric key pairs known as pre-keys help encrypt messages between communicating entities. The goal is to assure forward and backward secrecy even if one or more pre-keys are compromised. 
    \item The public components of the pre-keys are signed using the long-term identity key and communicated via the server to other communicating entities. The assumption is that only the account owner can create such signatures. 
    \item The private components of \(IK\) and the pre-keys are not known to anyone else, including the application provider server. Thus the communications are encrypted end-to-end. 
\end{enumerate}

\subsubsection{Mobile Applications}
Signal, WhatsApp, Element and Viber broadly adhere to the Signal protocol's Double Ratchet algorithm~\cite{dh_algorithm}, as shown in Table \ref{tab:summary-e2ee}. In essence, every communication sequence is encrypted with the pre-key listed against the intended recipient. The process of deriving secrets from previously held secrets is termed as ratcheting and is initialized with a shared secret. Sequence numbers help to decrypt out-of-order messages. The root of trust for each participant is the identity key they configured during account setup.

Wickr Me uses the Wickr secure messaging protocol~\cite{wickrme}, which pins key material to identifiers generated by devices, typically in the form of SHA-256 values. The application/device is the primary actor in this protocol and the basis of all trust relationships. The important primitives are the identity keys \(IK\), ephemeral asymmetric key pairs (similar to pre-keys) and a platform-specific message encryption key.  

Telegram is a messenger based on MTProto 2.0. Client and server share a device-specific 2048-bit permanent authorisation key created by a Diffie-Hellman (DH) key exchange~\cite{diffie1976new}. Communication can happen through a cloud server, or users can switch to secret chat, giving E2EE communication where short-lived keys are linked to the device-specific permanent authorisation key. 

\subsubsection{Desktop Clients}
Many E2EE apps offer a desktop client that can piggyback on the E2EE functionality of the phone or other primary device. These clients share some common characteristics: %  we include for this paper, there are essential steps that are common to all of them:
\begin{itemize}
    \item Every user needs to install the mobile application on a primary device which they control and have an account through it on the messenger service. 
    \item After installation, the desktop client for a  messaging application will typically generate its own identity key pair, distinct from that of the primary device. 
    \item The primary device and the desktop client authenticate each other. The primary device then tells the application server that the desktop client is trusted and can communicate as if it were the primary device. 
    \item The primary device retains a list of the linked companion devices. 
\end{itemize}

All the E2EE applications we investigated assume that account holders will be able to protect the private component of their identity key. The documentation does not typically give any advice as to how; many providers of messaging services consider endpoint security to be out-of-scope. This sits uncomfortably with the fact that mechanisms such as ratchets are designed to recover from transient device compromise by supporting forward and backward security properties. We elaborate on this further in our analysis.

\begin{table*}[t]
\centering
\footnotesize
\begin{tabular}{|l|l|l|l|}
\hline
        Applications & Protocol                                                                               & Primary Device (Phone) Parameters                                                                                                                                   & Desktop Client                                                                                                                                                                                   \\ \hline
Signal   & Signal                                                                                 & \begin{tabular}[c]{@{}l@{}}Curve25519 Key pair – Long term \\ Identity Key\\ Curve25519 Key pair – Pre-Keys\\ \end{tabular}                                  & \begin{tabular}[c]{@{}l@{}}Desktop ID authenticated by \\ primary device. \\ Can be used independently. \\ \end{tabular}                                         \\ \hline
WhatsApp & Signal                                                                                 & \begin{tabular}[c]{@{}l@{}}Curve25519 Key pair – \\ Long term Identity Key\\ Curve25519 Key pair – Pre-Keys\\\end{tabular}                      & \begin{tabular}[c]{@{}l@{}}Desktop ID authenticated by \\ primary device\\Can be used independently\end{tabular}  \\ \hline

Element  & \begin{tabular}[c]{@{}l@{}}Olm- \\ Double Ratchet \\ Implementation\end{tabular}       & \begin{tabular}[c]{@{}l@{}}Curve25519 Key pair – \\ Long term Identity Key\\ Curve25519 Key pair – Pre-Keys\\\end{tabular}                      & \begin{tabular}[c]{@{}l@{}}Desktop ID authenticated by \\ primary device. \\ Can be used independently. \\ \end{tabular} \\ \hline

Wickr Me & \begin{tabular}[c]{@{}l@{}}Wickr Secure \\ Messaging Protocol\end{tabular}             & \begin{tabular}[c]{@{}l@{}}Curve P521 Key pairs\\ SHA-256 Device Identifier \end{tabular}                                                                                  & \begin{tabular}[c]{@{}l@{}}Desktop ID authenticated \\ by primary device \\ Can be used independently. \end{tabular} \\ \hline

Viber    & \begin{tabular}[c]{@{}l@{}}Double \\ Ratchet Implementation\end{tabular}               & \begin{tabular}[c]{@{}l@{}}Curve25519 Key pair – \\ Long term Identity Key\\ \end{tabular}      & \begin{tabular}[c]{@{}l@{}}Desktop client authenticated \\ by primary device \\ Can be used independently.\\ \end{tabular}                                                                                   \\ \hline
Telegram & \begin{tabular}[c]{@{}l@{}}MTProto 2.0 – \\ Diffie \\ Hellman Implementation\end{tabular} & \begin{tabular}[c]{@{}l@{}}Cloud chat – 2048 bit permanent key\\ Secret Chat – \\ DH keys between communicating entities.\\ \end{tabular} & \begin{tabular}[c]{@{}l@{}}Desktop ID authenticated \\ by primary device \\ Can be used independently.\\ \end{tabular}                                         \\ \hline
\end{tabular}
\caption{Properties of Popular Messaging Applications}
\label{tab:summary-e2ee}
\end{table*}

\subsection{Threat Modeling}
Threat modelling is a key enabler for security \& privacy by design~\cite{cybok-ssl}. As a case in point, Microsoft's secure development lifecycle relies on it~\cite{howard2006security}. Its approach to systematic elicitation of security threats to the data flows between various system components is known as STRIDE. Some threats to privacy are distinct from those in STRIDE, and so a second methodology, LINDDUN, captures threats to privacy, such as the ability of an adversary to link actions to principals and thus defeat anonymity. We use both, in an attempt to capture the security and privacy threats at a reasonable level of granularity, and using industry-standard tools.

\subsubsection{STRIDE}

In STRIDE, the target system is modelled using Data flow Diagrams (DFDs) that capture the key components (processes and data stores), and the data flows between them. Trust boundaries mark which parts of the system are assumed to be free from adversarial interference. Data flows across trust boundaries and individual components are then evaluated for their susceptibility to six key threats: Spoofing, Tampering, Repudiation, Information disclosure, Denial of service, and Elevation of privilege (the capitalisation reflects the STRIDE acronym). These threats are modelled as the negation of five security properties: authentication, integrity, non-repudiation, confidentiality, availability and authorization.

\subsubsection{LINDDUN}

LINDDUN follows a similar approach to STRIDE but its focus is on threats to privacy. The system is once more modelled using DFDs, but in this case, the  individual DFD elements and data flows are evaluated for seven privacy threats: Linkability, Identifiability, Non-repudiation, Detectability, information Disclosure, content Unawareness and Noncompliance. These are once more the negation of seven privacy properties: unlinkability, anonymity/pseudonymity, plausible deniability, undetectability/unobservability, confidentiality, content awareness and policy/consent compliance.

\section{Methodology}
\label{methodology}

\subsection{Choice of E2EE Applications}
% We experiment with short-lived adversarial access to the desktop clients of some of the popular E2EE messaging applications. For our experiments,
We selected applications that are widely used and diverse in how they establish trust between primary and companion devices. They can be divided into two broad categories: those that rely on variants of the Signal protocol, and those that do not. For Signal-based apps, we study two implementations, Signal's and WhatsApp's; Signal's is open-source while WhatsApp's is not. As we will note later, there are noticeable differences. While not relying directly on the Signal protocol, Viber and Element both make use of its Double Ratchet algorithm. Viber differs from other implementations in the way the Root ID is shared between the primary and companion devices. We examine whether this affects the ability to protect against threats from short-lived adversarial access.
Element is noteworthy for being decentralized; it does not rely on a central communications server. We investigate if this has any bearing on trust establishment between companion and primary devices.
 
Two further messaging services, Wickr Me and Telegram, rely entirely on their own messaging protocols. Wickr documentation indicates that device-specific information is used in device enrolment, so we evaluate whether this is sufficient to prevent silent desktop cloning. Telegram uses a custom protocol that distinguishes between ``cloud'' chats and ``secret'' chats. The documentation does not discuss any measures for forward secrecy post-compromise.

\subsection{Creation of DFDs}
\begin{figure}[h]
  \centering
       \fbox{\includegraphics[width=0.9\columnwidth, height=2in]{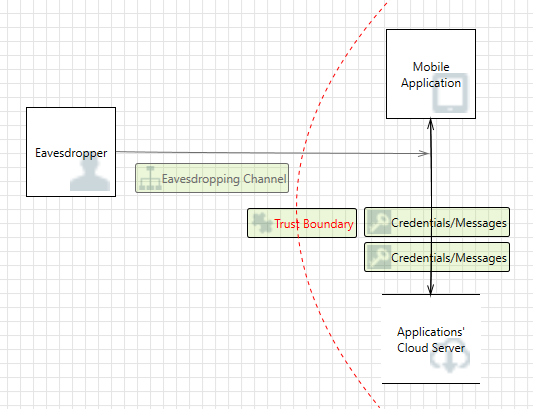}}
   \caption{DFD for Signal, WhatsApp, Element, Wickr Me, Viber, and Telegram mobile applications.}
   \label{fig:dfdm}
\end{figure}

We created DFDs (using Microsoft's Threat Modeling Tool) for each app before and after the addition of the desktop client, first by studying their documentation~\cite{signal_user, whatsapp, element, wickrme, telegram,viber} and second, through experiments (see Section~\ref{sec:experimental_setup}). 
For example, WhatsApp documentation explicitly states (page 25)~\cite{whatsapp} 
\begin{quote}
    ``\textit{WhatsApp defines end-to-end encryption as communications that remain encrypted from a device controlled by the sender to one controlled by the recipient, where no third parties, not even WhatsApp or the parent company Facebook, can access the content in between.}''
\end{quote} 
% We used one of Microsoft's threat modelling tools~\cite{mstmt} to prepare the DFDs for WhatsApp mobile application \textemdash{}
 %After configuring the desktop client of WhatsApp we investigate protections from short live adversarial access by mal-actors at one of the end points.
The \textcolor{red}{red} dotted line indicates the trust boundary that demarcates the security-critical artefacts. Figure~\ref{fig:dfdm} depicts the DFD at time \(t_{1}\) and space \(S_{1}\) (cf. Figure~\ref{fig:intro-threatmodels}) for the six mobile messaging applications. The private part of the \(IK\) and the pre-keys are inside the mobile device in which they were generated. Operation without a linked desktop client does not entail sharing them with any other device. There the designers consider that the secrets, being in the device, are under the control of their owners. The documentation advises that any device compromise should be reported and credentials revoked. So we place the security artefacts within the \textcolor{red}{red} trust boundary, keeping the eavesdropper outside.

\begin{figure}[ht]
  \centering
       \fbox{\includegraphics[width=0.95\columnwidth]{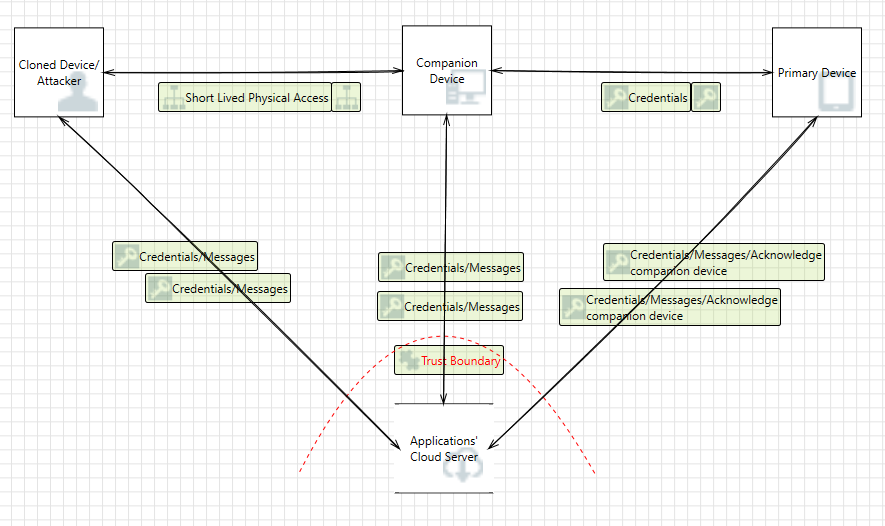}}
   \caption{DFD for Signal, WhatsApp, and Telegram desktop applications.}
   \label{fig:dfdsw}
\end{figure}

\begin{figure}[ht]
  \centering
       \fbox{\includegraphics[width=0.95\columnwidth]{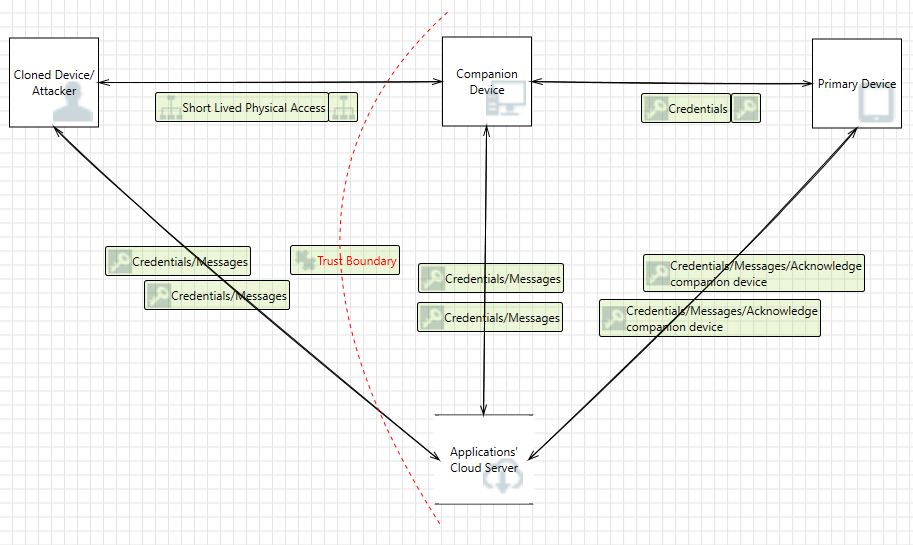}}
   \caption{DFD for Element, Wickr Me and Viber desktop applications.}
   \label{fig:dfdew}
\end{figure}

Figures~\ref{fig:dfdsw} and~\ref{fig:dfdew} show the DFD at time \(t_{2}\) and space \(S_{2}\) (cf. Figure~\ref{fig:intro-threatmodels}) where a desktop client has been added. The difference in the way companion devices are set up leads to two different DFDs. While all the desktop clients are first launched using the primary mobile device to which they are linked, there are then two separate cases. For Signal, WhatsApp and Telegram subsequent desktop clients can be launched without using the primary mobile device, but this is not the case for Viber, Element and WickrMe. 

The \textcolor{red}{red} dotted line in Figure \ref{fig:dfdsw} represents the trust boundary and thus the placement of security artefacts for Signal, WhatsApp and Telegram. The private part of \(IK\) and the pre-keys are within the realm of everyone with access to the desktop.% so they are indicated outside the trust boundary (i.e. \textcolor{red}{red} dotted line). 

Figure \ref{fig:dfdew} represents the placement of the security artefacts for Viber, Element and WickrMe. The trust boundary includes the private part of \(IK\) and the pre-keys. The reason is that every instance of a desktop client needs to be explicitly launched by the principal mobile device. As the keys cannot be copied, desktop clients cannot be launched independently. 

% A similar process combining the documentation along with practical configuration and experimentation is repeated to construct the DFDs for the other mobile applications and their desktop clients we consider for this paper.  

\subsection{Experimental Setup}\label{sec:experimental_setup}
The experiments were conducted between test accounts registered to phone numbers provided by pay-as-you-go SIM cards purchased specifically for these experiments. Account registration was performed using a Samsung Galaxy A21 smartphone and iPhone SE which were used to receive SMS messages required for registration. The desktop clients were installed through the following steps:
\begin{enumerate}
  \item \textbf{Hardware}. The desktop clients were installed on MacBook Pro laptops with 2 GHz Quad-Core Intel Core i5 and 16GB 3733 MHz LPDDR4X memory. 
%   \item \textbf{Installing the desktop clients}. The desktop clients of all the applications were installed on the MacBook Pro.
  \item \textbf{Firing up the legitimate desktop client}. We started a legitimate desktop version of the application through the required setup mechanism, for example, typically by scanning a \texttt{QR code}. 
  \item \textbf{Firing up the attacker's desktop client}. We performed a standard installation of the desktop client on an attacker's machine, configured with a second account, then copied the state from the victim's machine and placed it in the attacker's. Our goal was to evaluate if these systems protect against simple cloning attacks, so we copied state information from \texttt{$\sim$/Library/Application Support/} of the victim's machine to the same directory of the attacker's machine. The victim's machine and the attacker's machine were of the same specification. 
\end{enumerate}

\subsection{Testing for threats}
%\hl{This subsection and the next is again too focused on Signal. The testing for security threats and privacy threats should be part of the box and line process diagram I mention above. Again, here what you want to do is to be clear on what was done across the board (not just for Signal; all apps; it should all be the same so remove references to Signal). TM delta is currently discussed as the last step in the previous section but surely TM delta comes from the STRIDE and LINDDUN analysis. Related to this nowhere we say where we derived TM1 and TM2. This should be in the DFDs section above. We say we have the DFDs but they are not the threat models. This is the sort of thing ICSE reviewers will pick up on quickly. We need to be clear in that section that we used the DFDs to develop TM1 and (expected) TM2 using both STRIDE and LINDDUN. The subsections below then make sense as they are used to identify if TM2 holds or not. ANd if not how far is TM(exhibited) - as in derived from testing for each app - from TM2 (i.e., the delta).}

\subsubsection{STRIDE}
\begin{itemize}
    \item \textbf{Spoofing}. We performed a standard installation of the desktop client in an attacker's machine, then copied the state from the victim's machine to the attacker's.  
    \item \textbf{Tampering}. Our focus was on the endpoints rather than the network, so we did not experiment with altering message content while in transit.  
    \item \textbf{Repudiation}. While we set up the attacker's machine, we engaged in communication between the legitimate participant, the attacker and an innocent third party. This was repeated between the communicating entities to understand if third parties observe any difference when communicating with the victim and the attacker.  
   \item \textbf{Information Disclosure}. We used the cloned desktop across space and time to understand the implications of forward and backward secrecy. 
   \item \textbf{Denial of Service}. We tested whether the cloned machine throws the victim out of the network or allows the victim to continue sending and receiving messages even when the clone is in operation. 
   \item \textbf{Elevation of privilege}. This was tested by capturing the credentials using a \texttt{tls interceptor} from the rooted device. Then the victim's desktop client was de-linked from the primary device. Subsequently, the cloned desktop client was also de-linked. Then we used the captured credentials to restart the desktop client in the attacker's device. 
\end{itemize}

\subsubsection{LINDDUN}
% We experiment short-lived adversarial access to desktop clients and investigate their privacy consequences.
% With respect to LINDDUN
We focused on identifiability and linkability as information disclosure and non-repudiation versus repudiation were already evaluated in our STRIDE analysis. Since our focus was on endpoints, we did not perform an analysis for detectability. Non-compliance and unawareness are out of our scope for us as the registration processes for the messaging applications required primary device credentials without any room for opting out.
\begin{itemize}
    \item \textbf{Linkability}. -- The various artefacts from a victim were checked if they link potential entities connected to the victim. 
    \item \textbf{Identifiability}.  -- We checked if those artefacts revealed identifying information about the victims and indirect entities connected to the victim. 
\end{itemize}

We then analyzed the type(s) of data accessible through potential threats. We then modelled these data items as trees to depict the identifiability and linkability of an entity.
% Thereafter, to construct the attack trees for identified LINDDUN taxonomies, the respective data was arranged in a tree structure using root, leaf and child nodes. 

% \subsection{Interviews}

% \subsection{Ethics}

\section{Findings}
\label{findings}

A longstanding security goal in establishing a communication channel is to ensure that the only participants in the channel are legitimate ones. So desktop clients of E2EE messaging apps should be capable of identifying distinct participants in a channel, and thereby withstand  cloning attacks following short-lived adversarial access attacks. 
% \textbf{Calculating $TM\textsubscript{$\Delta$}$}.
We, therefore, subject the desktop client to short-lived adversarial access (\(TM_{2}\)) to elicit the threats based on the tests particular to STRIDE and LINDDUN. % As shown in the trust boundaries in Figures~\ref{fig:dfdsw} and~\ref{fig:dfdew},  
% The DFD for desktop clients in Figures \ref{fig:dfdsw} and \ref{fig:dfdew} are constructed and the trust boundary is delineated accordingly. 
% a standard installation in the mobile device generates a long term stable primary identity pertaining to a user which is then used to establish trust of the companion device with the remote server. The similarity in creating the root of trust in the mobile device is reflected in Figure \ref{fig:dfdm} where as resilience of the desktop clients against simple cloning attacks led to evolution of two DFDs as in Figures \ref{fig:dfdsw} and \ref{fig:dfdew}.
Table~\ref{tab:emerging_threats} shows the threats % denoted as $TM\textsubscript{$\Delta$}$ are those
which were not scoped while expanding from \(S_{1}\) to \(S_{2}\) between time \(t_{1}\) and \(t_{2}\).

% In our discussions we contrast the security properties of the mobile applications to explore if they have been expanded/adjusted to protect against the STRIDE and LINDDUN threat elements resulting out of \(TM_{2}\). 

\begin{table*}[!ht]
\centering
\footnotesize
\begin{tabular}{|c|ccccccccccccc|}
\hline
\multirow{2}{*}{\textbf{Applications}} & \multicolumn{13}{c|}{\textbf{Emerging Threats ($TM_\Delta$)}} \\ \cline{2-14} 
 & \multicolumn{1}{c|}{\textit{\textbf{S}}} & \multicolumn{1}{c|}{\textit{\textbf{T}}} & \multicolumn{1}{c|}{\textit{\textbf{R}}} & \multicolumn{1}{c|}{\textit{\textbf{I}}} & \multicolumn{1}{c|}{\textit{\textbf{D}}} & \multicolumn{1}{c|}{\textit{\textbf{E}}} & \multicolumn{1}{c|}{\textit{\textbf{L}}} & \multicolumn{1}{c|}{\textit{\textbf{I}}} & \multicolumn{1}{c|}{\textit{\textbf{N}}} & \multicolumn{1}{c|}{\textit{\textbf{D}}} & \multicolumn{1}{c|}{\textit{\textbf{D}}} & \multicolumn{1}{c|}{\textit{\textbf{U}}} & \textit{\textbf{N}} \\ \hline
\textbf{Signal} & \multicolumn{1}{c|}{\checkmark} & \multicolumn{1}{c|}{-} & \multicolumn{1}{c|}{\checkmark} & \multicolumn{1}{c|}{\checkmark} & \multicolumn{1}{c|}{$\times$} & \multicolumn{1}{c|}{\checkmark} & \multicolumn{1}{c|}{\checkmark} & \multicolumn{1}{c|}{\checkmark} & \multicolumn{1}{c|}{\checkmark} & \multicolumn{1}{c|}{-} & \multicolumn{1}{c|}{\checkmark} & \multicolumn{1}{c|}{-} & - \\ \hline
\textbf{Whatsapp} & \multicolumn{1}{c|}{\checkmark} & \multicolumn{1}{c|}{-} & \multicolumn{1}{c|}{\checkmark} & \multicolumn{1}{c|}{\checkmark} & \multicolumn{1}{c|}{$\times$} & \multicolumn{1}{c|}{$\times$} & \multicolumn{1}{c|}{\checkmark} & \multicolumn{1}{c|}{\checkmark} & \multicolumn{1}{c|}{\checkmark} & \multicolumn{1}{c|}{-} & \multicolumn{1}{c|}{\checkmark} & \multicolumn{1}{c|}{-} & - \\ \hline
\textbf{Element} & \multicolumn{1}{c|}{$\times$} & \multicolumn{1}{c|}{-} & \multicolumn{1}{c|}{$\times$} & \multicolumn{1}{c|}{\checkmark} & \multicolumn{1}{c|}{$\times$} & \multicolumn{1}{c|}{$\times$} & \multicolumn{1}{c|}{\checkmark} & \multicolumn{1}{c|}{$\times$} & \multicolumn{1}{c|}{$\times$} & \multicolumn{1}{c|}{-} & \multicolumn{1}{c|}{\checkmark} & \multicolumn{1}{c|}{-} & - \\ \hline
\textbf{Wickr Me} & \multicolumn{1}{c|}{$\times$} & \multicolumn{1}{c|}{-} & \multicolumn{1}{c|}{$\times$} & \multicolumn{1}{c|}{$\times$} & \multicolumn{1}{c|}{$\times$} & \multicolumn{1}{c|}{$\times$} & \multicolumn{1}{c|}{$\times$} & \multicolumn{1}{c|}{$\times$} & \multicolumn{1}{c|}{$\times$} & \multicolumn{1}{c|}{-} & \multicolumn{1}{c|}{$\times$} & \multicolumn{1}{c|}{-} & - \\ \hline
\textbf{Viber} & \multicolumn{1}{c|}{$\times$} & \multicolumn{1}{c|}{-} & \multicolumn{1}{c|}{$\times$} & \multicolumn{1}{c|}{$\times$} & \multicolumn{1}{c|}{$\times$} & \multicolumn{1}{c|}{$\times$} & \multicolumn{1}{c|}{$\times$} & \multicolumn{1}{c|}{$\times$} & \multicolumn{1}{c|}{$\times$} & \multicolumn{1}{c|}{-} & \multicolumn{1}{c|}{$\times$} & \multicolumn{1}{c|}{-} & - \\ \hline
\textbf{Telegram} & \multicolumn{1}{c|}{\checkmark} & \multicolumn{1}{c|}{-} & \multicolumn{1}{c|}{\checkmark} & \multicolumn{1}{c|}{\checkmark} & \multicolumn{1}{c|}{$\times$} & \multicolumn{1}{c|}{$\times$} & \multicolumn{1}{c|}{\checkmark} & \multicolumn{1}{c|}{\checkmark} & \multicolumn{1}{c|}{\checkmark} & \multicolumn{1}{c|}{-} & \multicolumn{1}{c|}{\checkmark} & \multicolumn{1}{c|}{-} & - \\ \hline
\end{tabular}
\caption{($TM_\Delta$) based on STRIDE and LINDDUN threat models. In this table, (-) indicates not being tested; ($\times$) indicates attack not possible; and (\checkmark) indicates attack is possible.}
\label{tab:emerging_threats}
\end{table*}

\subsection{Signal Messenger} 
Signal messenger assumes that only an eavesdropper can be, or attempt to be, the adversary and implements its authentication and key-sharing mechanisms accordingly. Against other adversaries, the expectation is that the user would replace the device or account. This may have been appropriate before companion devices were supported, as private keys never left the primary device. However, our experiments show that such assumptions crumble when potential adversaries reside within the trust boundary and adversarial short-lived access can go undetected. 

\paragraph{Masquerading as the victim} An attacker can simply replace the configuration files of a standard Signal desktop installation with the version stolen from a victim's machine. The specifics of the attacker machine do not influence the success of the attack. Signal desktop uses an encrypted SQLite database to store Signal authentication credentials: login and device password (the \texttt{uuid\_id} and \texttt{password} fields),  received messages, and pre-keys. The relevant files  are named \texttt{config.json} and \texttt{databases/Databases.db}. The database decryption key is stored in plaintext in the parent directory in \texttt{config.json}. For a knowledgeable adversary, the database is effectively stored in plaintext\footnote{This design decision has been discussed on multiple occasions on the Signal Community Forum and GitHub issue tracker since 2017.}. The collapse of the assumption of an \emph{only eavesdropper} threat model leads to the violation of other security properties with privacy consequences. 

When the Signal application is re-installed on a device, whether due to device compromise or for other reasons such as a new phone, the user identity key changes along with the pre-keys. This is notified to all their contacts, alerting them to the change and enabling them to verify new keys out-of-band if they wish. However, a comparison between the legitimate desktop version and a cloned desktop shows the same keys against the sequence number of the pre-keys. This brings home that the DH ratchet is not effective at rendering the cloned version obsolete after the existing key material is exhausted.

The mobile application assumes that the security protocol mechanisms will prevent an eavesdropper from capturing enough state information to masquerade as the victim -- there are message sequence numbers to prevent an eavesdropper from simply replaying past messages. In such a case, improper communication will be detected automatically. There is no such protection against an adversary with short-lived access to the desktop client. There are still overt symptoms that could be noticed by an alert victim. For example, the attacker's Signal desktop may work with delays or messages will be dropped when the victim's mirrored desktop installation is actively online. In none of our experiments, the active desktop thrashed visibly between the original and the clone. Our observation was that such behaviour was dependent on whether the session was established by the victim, the attacker or the communicating second party. On subsequent examination, we noted that the Signal protocol documentation~\cite{dh_algorithm} advises setting a time limit on the retention of skipped message keys in order to protect against an eavesdropper.

\paragraph{Compromising Forward Secrecy} The desktop client state information contains private pre-key material that will let the attacker break forward secrecy for the Signal account to which it is linked. Copying it results in the attack computer's Signal installation exactly mirroring the victim's installation. The attacker instance contains all historical messages from the victim installation and is able to receive and decrypt all future incoming messages as well as send encrypted messages that appear to be from the victim. The mobile device and legitimate desktop client do not have separate ratchets. 

We tested de-linking the legitimate Signal desktop client from the phone, which initially made the cloned desktop unable to send or receive messages. Since the database decryption key was easily available, we loaded the signal app into an adb-emulated Android device running on Ubuntu. We used run time level hooks provided by Frida~\cite{frida} to develop a \texttt{tls-interceptor} app. Then loaded the \texttt{tls interceptor} app into the Android device to capture the traffic. We were able to capture the primary device ID and password using the \texttt{tls-interceptor} and use them to update and reconnect the cloned (but de-linked) desktop client. We were then able to send and receive messages on the de-linked desktop client using the primary credentials. The details of the attacker machine do not appear to affect the success of this attack. This experiment was performed only for Signal as the database decryption keys were not so readily identifiable in the material cloned for other desktop clients.

\begin{figure}[ht]
  \centering
       \fbox{\includegraphics[width=0.95\columnwidth,height=1.75in]{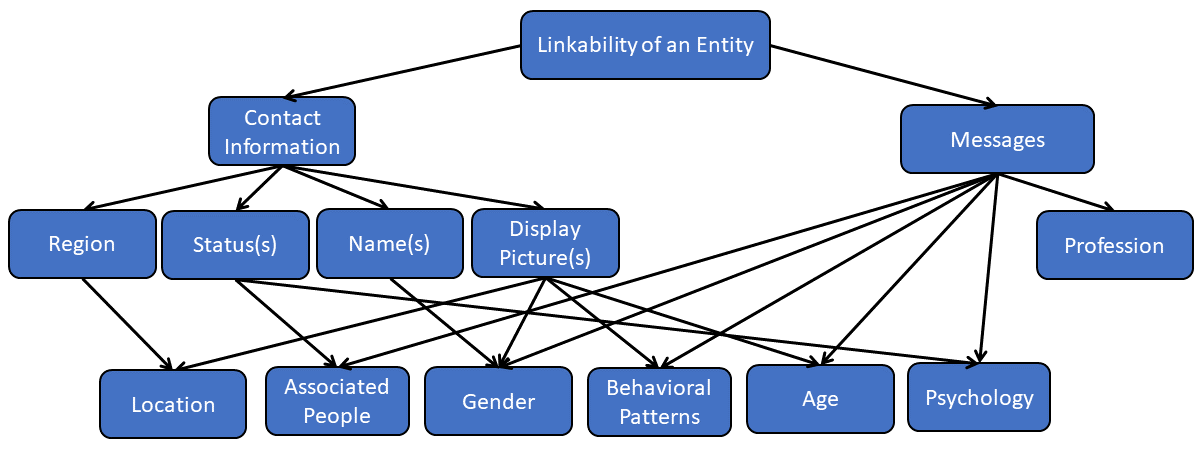}}
   \caption{Linkability of an Entity due to cloning of a device}
   \label{fig:attacktree-linkability}
\end{figure}

\begin{figure}[ht]
  \centering
       \fbox{\includegraphics[width=0.95\columnwidth,height=1.75in]{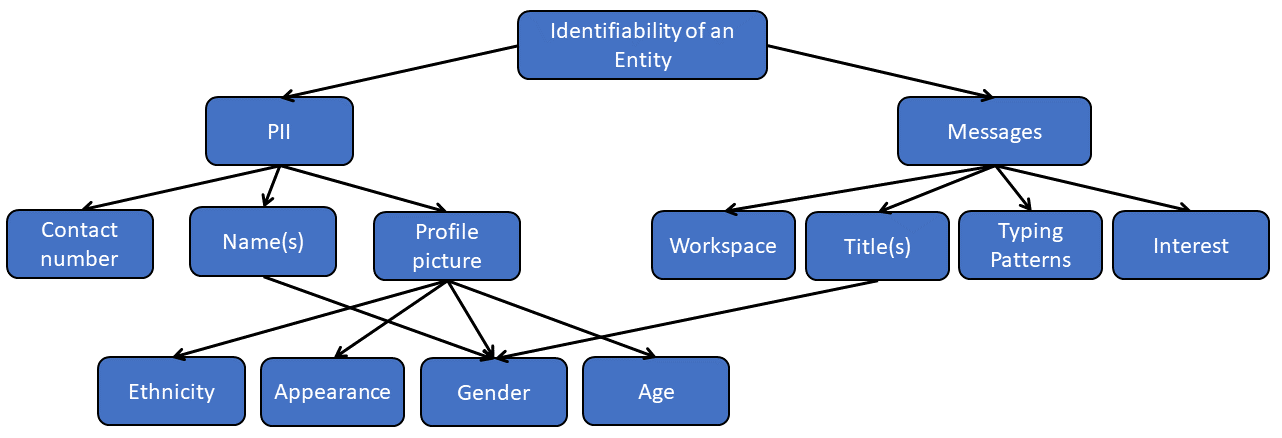}}
   \caption{Identifiability of an Entity due to cloning of a device}
   \label{fig:attacktree-identifiability}
\end{figure}

\paragraph{Un-scoped Threats, Security and Privacy Consequences}$TM\textsubscript{$\Delta$}$ for Signal desktop client reveals that it was not scoped for protection against \emph{spoofing, repudiation, information disclosure, denial of service and elevation of privilege} with respect to STRIDE and \emph{linkability and identifiability} for LINDDUN.
% The DFD elements that an adversary duplicates through short-lived access to the victim's desktop has privacy consequences. We take those elements to construct the attack trees pertaining to LINDDUN - linkability taxonomy
As shown by the example tree in Figure \ref{fig:attacktree-linkability}, the messages that an adversary has access to can be used to infer the inter-relationship between the contacts of the victim and with the victim. This can lead to secondary and tertiary linkability between the victim and entities through their direct contacts. Furthermore, as shown in Figure~\ref{fig:attacktree-identifiability}, a direct consequence of compromised DFD elements is that the attacker has further access to personally identifiable information and can identify the subjects and pair them with attributes. % For example their behavioral patterns their age and profile them at great depth. The identifiability attack tree subsequent to linkability is expressed as Figure \ref{fig:attacktree-identifiability}. 

\begin{finding}
  For the Signal messenger desktop, it is difficult to detect short-lived adversarial access and recover from a compromise. Access to the database decryption keys can render de-linking inconsequential. Adhering to the mobile-app threat model of external eavesdroppers betrays the reality that adversaries with transient access to linked devices might abuse this and escalate it to persistent access to the account. 
\end{finding}

\subsection{WhatsApp}
WhatsApp desktop clients consider malicious entities other than an eavesdropper. However, our experiments show that their protection mechanisms were not very effective against such an adversary. An attacker with short-lived access can steal the credentials and communicate as the victim. However unlike Signal the key is not stored in plain text.
This has a bearing on re-configuring a stolen desktop once it is de-linked from the primary device. Another improvement over Signal is that all companion devices have an expiry date. WhatsApp also alerts the primary device to new companion devices, as well as de-linking any existing companion devices in use. While spoofing is possible as in Signal, these design changes can limit the consequences. 

Messages are still synchronized across devices, thus compromising confidentiality. An attacker is able to send and receive messages as the victim. This breaks forward secrecy and the recipients are not able to tell whether the messages are from an attacker or the victim.

Though there is a mitigation in the form of expiry dates, there are further privacy consequences for linkability and identifiability. 

$TM\textsubscript{$\Delta$}$ for the WhatsApp desktop client reveals that it was not scoped for protection against \emph{spoofing, repudiation, information disclosure} with respect to STRIDE and \emph{linkability and identifiability} for LINDDUN. However, the WhatsApp desktop client was scoped for protection against \emph{elevation of privilege} and \emph{denial of service} through short-lived adversarial access. The scoping for \emph{spoofing} is a marginal improvement over the Signal desktop with the provision of expiry dates and alert messages for companion devices. 

\begin{finding}
  Short-lived adversarial access does not reveal the database keys which, coupled with the alerts and expiry dates, protects the WhatsApp desktop user. The desktop client threat model considers that legitimate insiders can turn malicious but makes strong assumptions about users' ability to note and act on warnings and protect themselves. 
\end{finding}

\subsection{Viber}   
Our experiments show that Viber clones exited as soon as they were launched. 
This is because Viber explicitly pins the primary identity in the companion devices \textemdash configuring the companion device transfers the identity key pair to it. This is significant for both the mobile application and the desktop client. Any device willing to authenticate as a legitimate client needs to be explicitly authorized by the primary device. Though the state is stored in \texttt{$\sim$/Library/Application Support/Viber}, using this state requires the explicit transfer of the primary identity key by the legitimate owner. 

When communicating entities are authenticated by each other without any interference from a man-in-the-middle, the trusted session is identified with a green lock. Any other session is marked with a red lock. 

For Viber, $TM\textsubscript{$\Delta$}$ reveals that it was scoped for protection against \emph{spoofing, repudiation, information disclosure, denial of service and elevation of privilege} with respect to STRIDE and \emph{linkability and identifiability} for LINDDUN.  

\begin{finding}
Viber appears well-scoped for the dynamic nature of the threat landscape by explicitly mandating companion devices through a transfer of the mobile device's primary ID. This eliminates the responsibility of users to detect cloning attacks and/or communications by the attacker and engage in the subsequent recovery.   
\end{finding}

\subsection{Wickr Me} Cloning attacks were not possible for Wickr Me even when the victim's state was set to 'remember password'. This implies that Wickr Me considered malicious participants beyond just eavesdroppers. Their mobile messaging application verifies the association of identity with its identity key pair and ephemeral key pairs. The association between the identity key pair and the identity is managed by the Wickr app and is pinned to the device, making it difficult for an attacker to authenticate as a victim. To protect from eavesdropping, Wickr Me encrypts server requests using a rotated shared secret using AES 256 in CFB mode which is tunnelled inside TLS. The security property of pinning the identity and key pair with the device also appears to be extended to the desktop client. 

$TM\textsubscript{$\Delta$}$ for Wickr Me reveals that it was scoped for protection against \emph{spoofing, repudiation, information disclosure, denial of service and elevation of privilege} with respect to STRIDE and \emph{linkability and identifiability} for LINDDUN.  

\begin{finding}
  Wickr Me is distinct in the way it ties the device to an instance of companion device identity establishment \textemdash{} a cloned instance of a Wickr Me account will not be able to masquerade as the victim. Such implementations acknowledge the changing nature of the threats from actors with access to the devices, and the desirability of robust and verifiable bindings between cryptographic keys and real-world entities.
\end{finding}

\subsection{Element} The desktop client considers malicious participants other than eavesdroppers to a greater extent than WhatsApp, but our cloning attacks still allow an attacker to find out who communicated with whom and when. While we moved the victim state to the attacker machine, the attacker was able to fire up the desktop client. The attacker was not able to send and receive messages and could not connect to the server, but they could see the user names of the entities with whom the victim communicated and when. 

The mobile version of Element generates a secret key for every user of a container particular to a device, and we believe they extended this to their desktop version. The Matrix documentation states that they generate keys per device and not per user and that keys are never exported~\cite{specsmatrix}. The keys cannot, therefore, be stolen and replicated in another device through a simple cloning attack. Thus, though cloning is possible in Element, it does not compromise forward and backward secrecy. However, the ability to see who sent messages to whom and when could still lead to linkability between the victim and their contacts. 

In the case of Element, $TM\textsubscript{$\Delta$}$ reveals that the desktop client was scoped for protection against \emph{spoofing, repudiation, denial of service and elevation of privilege} with respect to STRIDE and \emph{identifiability} for LINDDUN. Element desktop client also appears to be scoped for \emph{information disclosure} for forward and backward secrecy but not for \emph{linkability} with respect to LINDDUN.
 \begin{finding}
    An attacker cannot compromise forward secrecy in Element through short-lived adversarial access. Element considers the threat of malicious insider access, and blocks access to message contents, but leaves users vulnerable to traffic analysis: a cloning attack still reveals the identity of communicating entities. 
 \end{finding}

\subsection{Telegram}
The mobile application has a cloud-based chat and an end-to-end secure chat, using MTProto 2.0. The protection primitives assume that a user is in control of the device. This assumption means that the user in control of the device can only authenticate with the long-term shared identity key. However, the desktop client's state information can be cloned through short-lived access and thus spoofed by an adversary. 
It is difficult for a recipient to distinguish between a legitimate sender and an attacker using a cloned account. 

In the case of Telegram secret chats, message exchanges are not synchronized across legitimate and cloned devices. For secret chats, the client key pairs are replenished after every 100 messages or after being in use for more than a week. This is to prevent any compromise of forward secrecy. Participants in a secret chat can initiate key generation if and when they detect any compromise of their keys. However, attackers can also initiate secret chats with the contacts of the victim without the contacts or the victim being able to detect them. 
The disclosure of contact information leads to inferences about the contacts of the victims and leads to identifying sensitive information. % The attack tree for linkability and identifiability are as as Figures  \ref{fig:attacktree-linkability} and \ref{fig:attacktree-identifiability}. 

$TM\textsubscript{$\Delta$}$ reveals that Telegram Desktop client was not scoped for \emph{spoofing, repudiation, information disclosure} and \emph{denial of service} with respect to STRIDE and \emph{linkability and identifiability} for LINDDUN. The scoping for \emph{spoofing} was a marginal improvement over Signal Desktop with the provision of multi-factor authentication and passwords (as well as the option to set an automatic logout after a period of time). 

\begin{finding}
  Telegram desktop clients can be cloned and detection of a compromise is non-trivial for a victim. The ability to set expiry dates considers malicious insiders to some extent but leaves users to protect themselves. The adoption of the \emph{eavesdropper-only} threat model in a context where access to account state information is easier, leaves users vulnerable to cloning attacks.  
\end{finding}

%\textemdash{} we go through the documentation of the messaging applications to understand their original threat model. The understanding is used to construct the DFD in Figure \ref{fig:dfdm} and its corresponding threat model \(TM_{1}\) where the protection mechanisms address an eavesdropper. as Figures \ref{fig:dfdsw} and \ref{fig:dfdew} respectively

\section{Threat modelling to align trust boundaries with administrative boundaries} 

\begin{figure*}[!t]
     \centering
     \begin{subfigure}[b]{0.3\textwidth}
         \centering
         \includegraphics[width=0.9\textwidth]{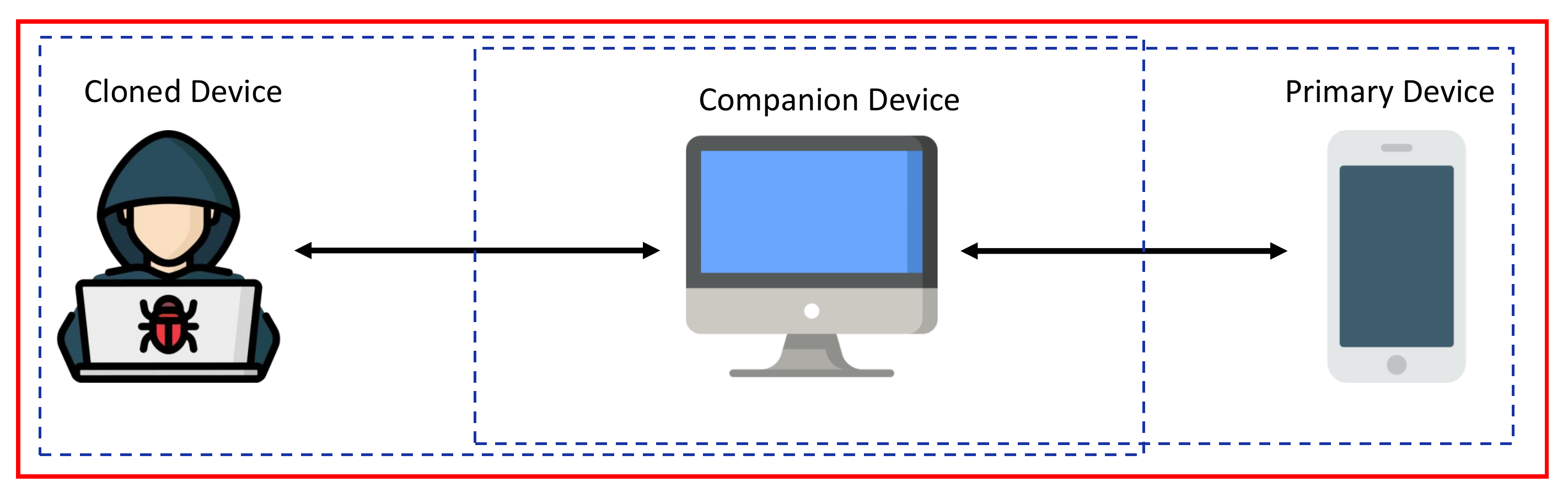}
         \caption{Mal-actors within the trust boundary}
         \label{fig:swt}
     \end{subfigure}
     \hfill
     \begin{subfigure}[b]{0.3\textwidth}
         \centering
         \includegraphics[width=0.9\textwidth]{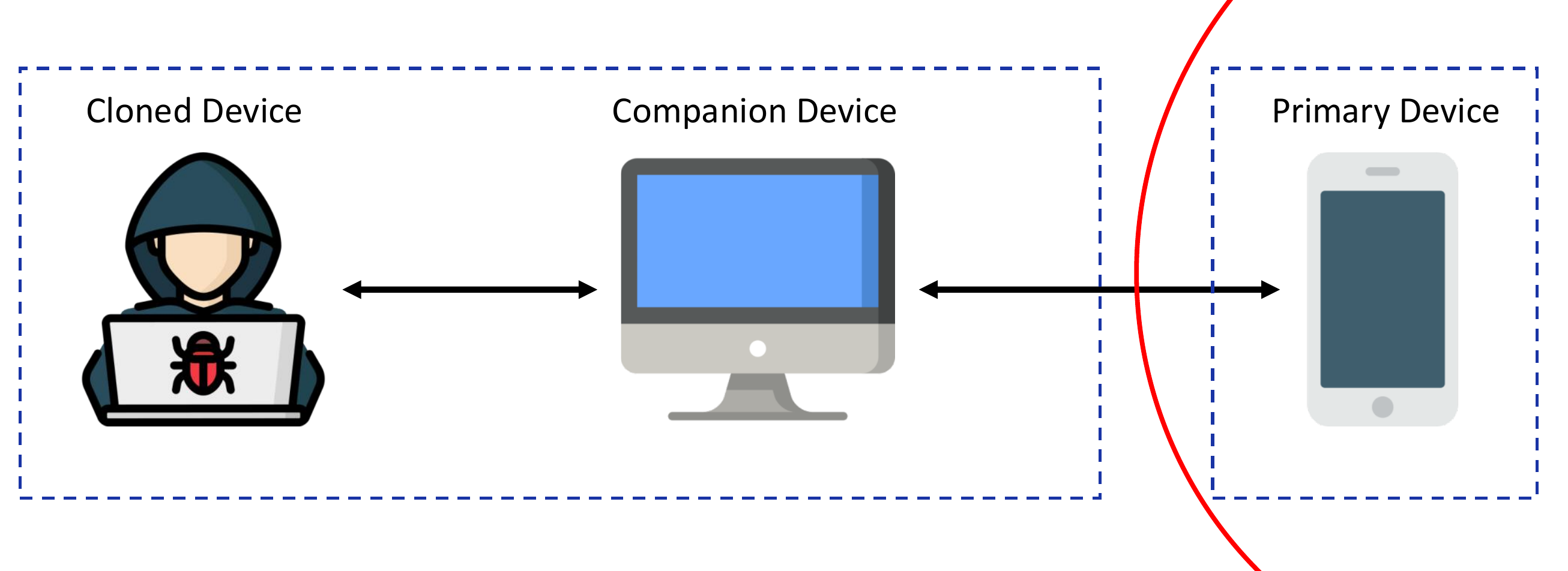}
         \caption{Primary device only in trust boundary}
         \label{fig:m}
     \end{subfigure}
     \hfill
     \begin{subfigure}[b]{0.3\textwidth}
         \centering
         \includegraphics[width=0.9\textwidth]{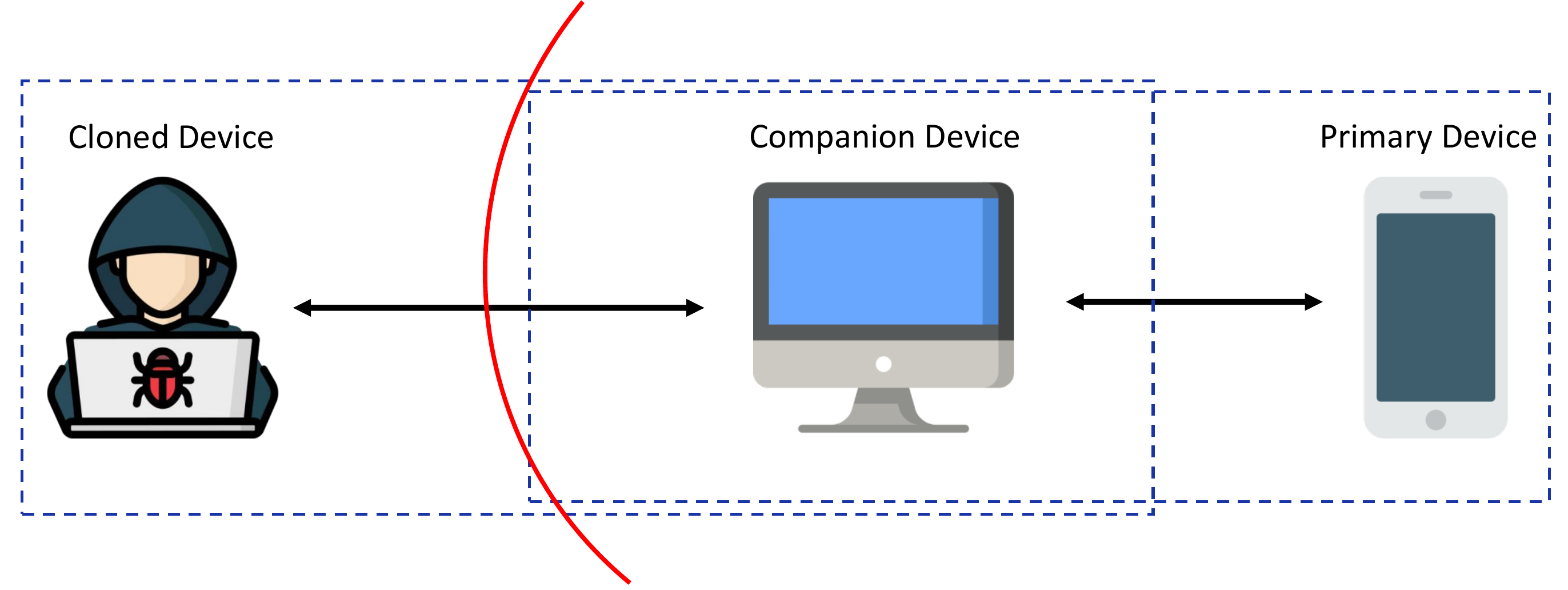}
         \caption{Mal-actors outside the trust boundary}
         \label{fig:vew}
     \end{subfigure}
        \caption{Placement of trust boundary (red line) with respect to administrative boundary (blue dotted line).}
        \label{fig:placement}
\end{figure*}

Distinct individual or collective entities function within a logical space (personal space, departments or organizations). We refer to these as \emph{administrative boundaries}. System designers evaluate threats within and across these administrative boundaries, specify security policies to counter the threats and provide appropriate controls to implement them. The artefacts protected by controls are placed within a \emph{trust boundary} in order to mitigate particular threats. The key question is the extent to which trust and administrative boundaries should align. Delineating trust boundaries too restrictively will impose a heavy burden on legitimate users e.g., frequent re-authentication. On the other hand, not scoping the trust boundary suitably can lead to bad actors within the administrative boundary being able to compromise security and privacy. And as discussed above, threat models can change \textemdash{} new threats can emerge to an existing application~\cite{omoronyia2013engineering,sgandurra2016evolution}. And additional features or extensions can change the administrative boundary leading to the emergence of new threats. This phenomenon is observable in our case study.

Mobile phones function within a culture where sharing of devices is not common. The threat model is focused more on external entities like eavesdroppers and middleboxes. System designers accordingly defined the security policies for their E2EE services to focus on the mobile device and implemented controls to protect against external threat actors. This is reflected in the trust boundary of the DFD in Figure~\ref{fig:dfdm} pertaining to  $TM\textsubscript{$1$}$, which contains the mobile phone.  

Then when we see desktop clients added to messaging applications, $TM\textsubscript{$2$}$ represents scenarios where there is a shift in the administrative boundary \textemdash{} external participants in official or domestic settings have easier access to the desktop clients, and normally legitimate insiders can turn malicious. 
Our investigation reveals distinct positions of the trust boundary depending on how companion devices can be fired up pertaining to their corresponding user account. These are captured in Figure \ref{fig:placement}. Signal messenger, WhatsApp and Telegram permit companion devices to be set up without the primary device; they can be cloned from an initial companion device with key material loaded by the primary device \textemdash{} allowing adversaries with short-lived adversarial access to a genuine companion device to clone it. Figure~\ref{fig:swt} shows that this is because the trust boundary includes legitimate insiders who can turn malicious. Compare the scenario where the trust boundary contains only the mobile phone as in Figure~\ref{fig:m}; this is secure but would require frequent authorization by the primary device. 

On the other hand, the security controls for Viber, Element and Wickr Me require all companion devices to be fired up by the primary device. The trust boundary as depicted in Figure~\ref{fig:vew} includes the primary device and only those desktop clients explicitly fired by it. This excludes any potentially malicious insider with short-lived access to a client. 

The question is how security engineering can rescope the trust boundary with changes in the administrative boundary as applications evolve over space and time. There have been several studies on the challenges of threat modelling in agile development environments~\cite{galvez2018odyssey} which acknowledge the emergence and disappearance of threats. A related study~\cite{cruzes2018challenges} reports operational challenges with threat modelling in agile environments and recommends including security experts on developer teams. The importance of experts is further supported by Assal~\etal{} reported that the absence of security knowledge is a significant blocker in the adoption of secure coding practices by developer teams~\cite{assal2018}. 

Our investigation suggests that re-evaluation of trust boundaries in the light of $TM\textsubscript{$\Delta$}$ need not be a highly resource-intensive task. For instance, it might build on lightweight interventions proposed in previous practice studies such as Weir~\etal{}~\cite{weir2019interventions}. They designed low-cost practical support for development teams \textemdash{} \emph{Developer Security Essentials}. They propose that threat assessment could be low-cost and easy to implement with the aid of a suitable facilitator. A specific output of the threat assessment can be the evaluation of $TM\textsubscript{$\Delta$}$ with respect to specific threat taxonomies.     

\subsection{Scoping `often' to protect human rights}

Appropriate scoping of the trust boundary with respect to the administrative boundary has implications for the human rights of direct users and implicated data subjects of E2EE messaging. Given the push in multiple jurisdictions for embedding content scanning tools in E2EE messaging applications, using arguments around protecting children and preventing terrorist radicalisation, there is an incentive for some principals to obscure trust boundaries. Examples include police and intelligence agencies claiming that end-to-end encryption is not compromised by mandating government scanning of message content before it is encrypted and sent, or after it is decrypted and received. Clarity on trust boundaries and administrative boundaries may help legislators see through such arguments and come to more appropriate decisions.

We draw on the first public evaluation~\cite{peersman2023towards} of some prototype client-side scanning tools to highlight the importance of appropriate scoping of the trust boundary in technologies that claim to protect children without violating individual privacy. 

\paragraph{Threats resulting from expanded memory scanning} A key consideration is whether content scanning tools can scan everything in the device where they are deployed. If they can, then mechanisms mandated to undermine the privacy offered by one messaging app will destroy the security and privacy of all other data on the device as well. The threat modelling approach described here will show that tools placed at the operating system layer, and with broad administrative privileges, situate themselves within the trust boundary of many security-critical artefacts. 

The evaluation document~\cite{peersman2023towards} brings out the varying degrees to which pervasive memory scanning is performed by the candidate surveillance tools. Some allow pervasive scanning while others leave some discretion to the user. Even so, there will be serious issues around whether defaults are safe, and in line with citizens' expectations, and indeed with human rights law. Tools that undermine platform security rather than just circumvent the encryption mechanisms of an app with which they are bundled can lead to pervasive surveillance and abuse.  

\paragraph{Threats due to embedding the tools within other applications} Content scanning tools can be placed within E2EE messaging applications or other applications, undermining their security \& privacy controls. This has been flagged by the evaluation report~\cite{peersman2023towards}. For example, video conferencing and gaming platforms might be mandated to embed such tools. Threat modelling needs to consider cases where such users' security or privacy protection might be turned off remotely, whether coercively or surreptitiously or both, leaving the application re-purposed and the end user vulnerable. Appropriate scoping of the trust boundary will need to take into account the complex and diverse underlying memory protections offered by Windows, Android and iOS~\cite{abelson2021bugs}. It is quite possible that an app with a mandated government backdoor might be a systemic surveillance and security threat on some platforms but not on others. This might lead to interesting policy externalities. For example, if a mandated backdoor in WhatsApp threatens Signal too on Windows devices but not on Apple devices, then users might either abandon WhatsApp or abandon Windows. Microsoft might then be motivated either to harden Windows or to lobby harder against the surveillance mandate.

%\emph{(2) Conflicting threat perceptions and priorities.} The content scanning tools depend on the security \& privacy settings of the E2EE messaging applications to prevent any ab
%\ \\

% \begin{finding}
%  Agile environments can include the calculation of $TM\textsubscript{$\Delta$}$ as part of every \emph{epic} as an \emph{investigation}. Security and usable security experts can come in for the grooming sessions to help developers build the \emph{enabler} and/or \emph{story} pertaining to the security controls to address threats pertaining to the particular \emph{epic} feature scope. This exercise can iterate for every \emph{epic} feature scope till the stable final release of an application. 
% \end{finding}

% The trust boundary for Viber, Element and WickrMe includes the companion devices within the trust boundary as in Figure \ref{fig:dfdsw}. The resilience of the existing security policies and corresponding controls come under a test. 

%\input{interviews}
\section{Discussion}
\paragraph{Reconciliation of security requirements across components with shared state}
The E2EE messaging applications were initially designed for mobile phones and the desktop clients followed later on \textemdash{} mobile phone application continued to be the root of trust for the desktop clients. A pertinent issue is if `trust' in the security of the mobile application is enough to trust the security of the desktop clients. On the other hand, can the compromise of the desktop client lower the security of the mobile application account as well? 

Setting up a desktop client involves sharing credentials between the primary device and the secondary device \textemdash{} the credential is the shared state between the primary and companion device.  As our analysis shows, for some of the desktop clients, the shared system state is open to compromise due to short-lived adversarial access.

Research in secure systems development has considered the question of when the security of the components is sufficient to trust the larger system (i.e., the composability problem)~\cite{ross2016systems}. Threat modelling of individual system components can help identify the shared state between them and analyze the consequences of such sharing. If sharing is inescapable then security (of the larger system) is perhaps incumbent on the administration of the shared components. In the context of security of open distributed processing, Bull~\etal{} suggest independent administration of the components~\cite{bull1992towards}. Gong~\etal{} proposed a model for autonomous administration of shared state for secure inter-operation of systems~\cite{gong1996computational}. Minimal sharing of state information has been a longstanding security principle, e.g., \emph{least common mechanism} in the Saltzer \& Schroeder principles~\cite{saltzer1975protection}. 

% In the context of a compromised desktop client affecting the mobile application account, it is relevant to mention that Wickr Me \& Element allow the desktop client to suspend and sign out the mobile application respectively. WhatsApp and Viber does not allow the actions in a desktop client to affect their mobile application user.  

\paragraph{Safe Defaults} 
Systems are designed with clear delineation of the participants in the system. We learn from our investigation of the desktop clients of the messaging applications that some of them assume that these participants have fixed behaviour which does not change across space and time. % We also observe the contrary in case of some others.
The manner in which applications respond to the change in the behaviour of the participants determines the security of the applications. \emph{Fail-safe defaults} have been a long-standing principle too~\cite{saltzer1975protection}. 

Developers, administrators and end users find it challenging to make fail-safe design and usage decisions~\cite{jaeger2021toward}. There are some ways to facilitate such decisions: prompting and nudging developers when committing code may help~\cite{chenprompts}. There have been suggestions of contextualizing the principles for application developers; see for example Neumann's discussion of the limits of willpower, and what principles can achieve~\cite{neumann2018fundamental}. Overall, there appears to be a consensus that we still need more work on ways to make fail-safe decisions~\cite{jaeger2021toward}. 

% In our extant case the threats arise from whether or not companion devices are included within trust boundaries with respect to $TM\textsubscript{$2$}$. The extent of threats scoped by the desktop clients of six messaging applications are captured as $TM\textsubscript{$\Delta$}$. As a technical exposition of safe defaults we can cite Viber, Wickr Me (and Element to a great extent), where handing over trust to companion devices is explicit and minimal. 

% \paragraph{Was threat modeling across space and time practiced?} \hl{This is where we describe that we reached out but got no response or only one but off-the-record so it can't be discussed.}

%Threats (namely, short-lived adversarial access) to the desktop clients highlight the difference in their security requirements compared to the mobile application. 
\section{Related Work}\label{sec:related}

There is a sizable body of work investigating and demonstrating vulnerabilities in E2EE messaging protocols and their implementations. Prior work has focused particularly on Telegram due to its use of a custom protocol, and researchers have demonstrated numerous attacks over the years on both the MTProto protocol and its implementation \cite{jakobsen2015practical, albrecht2022four}. Most recently, the Matrix protocol (which Element uses) suffered several severe vulnerabilities in which a malicious actor was able to break the confidentiality of communications using a compromised Matrix server \cite{albrecht2022matrix}.

Research has also demonstrated the technical feasibility of device cloning as a means of accessing user accounts on E2EE messaging platforms. Cremers~\etal{}~\cite{cremers2020clone} investigated practical post-compromise security measures in the mobile clients and some desktop clients of the main E2EE messaging platforms, finding that almost all clients studied are vulnerable to some degree of cloning attack. The authors identify challenges in consistently maintaining state across devices as the primary practical reason why these platforms do not attempt to prevent cloning: a post-compromise security mechanism can run the risk of being overly strict (cf. Figure~\ref{fig:m} and lock legitimate users out of their devices if synchronization fails, leading to usability concerns. However, our experiments show that the desktop clients of Viber, WickrMe and Element do better in detecting cloning attempts than their mobile counterparts~\cite{cremers2020clone}. Cisco's threat intelligence unit further experimented with cloning a handful of E2EE desktop clients (specifically, Signal, WhatsApp, and Telegram) in the wake of the 2018 `Telegrab' malware strain that took advantage of Telegram's vulnerability to desktop cloning~\cite{cisco_cloning2018}. Similarly to Cremers et al., they found that all three desktop clients studied allowed a cloned client to masquerade as legitimate, sending and receiving messages with little to no indication to the original user.

In this paper, we contribute to this area by analyzing the problem from a threat modelling perspective. We demonstrate that the problem arises from a lack of consideration of threats across space and time. The evolution of applications is a reality as is the addition of new features. However, rescoping the trust boundary as both the application and the threats evolve is critical. Our analysis of the delineation of trust and administrative boundaries provides a basis for maintaining security as systems and their environments evolve and at the tactical level for better administration of shared components, including the design of safe defaults.
% .  systematically analysing the consequences of desktop clients' vulnerability to cloning attacks against notable security and privacy threat elicitation frameworks. We present a taxonomy of threats against six of the most widely used E2EE messaging services as these systems transitioned from mobile-only to multi-platform. We conclude with a discussion of design considerations with the potential to mitigate cloning vulnerability.

\section{Conclusion}

Software design can involve a tussle between functional requirements and their security implications, where functionality all too often wins.  In an investigation of messaging clients, we have observed two contrasting ways in which they deal with the threats they face. One set of applications anticipated the possibility of client cloning and implemented systems-level mitigation strategies, while another put the onus squarely on the user to prevent device compromise in the first place. While most of the desktop clients are based on similar cryptographic primitives, their diversity is a reflection of their varying perceptions of likely attackers. Designers with a more realistic appreciation of the threat produced more robust systems. The larger lesson here is system maintainers have to keep their threat model up to date, especially if their product is successful and acquires millions of users, or they will be left behind and a mismatch between their threat model and reality may leave users exposed to attacks. 

We argue for nuanced and in-depth modelling of attackers in appropriate contexts as integral to the software development lifecycle. Application features evolve -- existing features are deprecated or updated and new ones added -- and so must threat models. The model of the attacker cannot be independent of the model, and indeed of measurement, of system users. There are users of shared devices and or managed devices. The administrative boundaries expand with the participation of occasional adversarial users such as border \& customs officials, nosy bosses and abusive intimate partners. The growing diversity of entities within the administrative domain exposes users to previously unscoped threats.  

Our suggestion for application designers is to not only do threat modelling little and often but also to pay close attention to $TM\textsubscript{$\Delta$}$ when doing so. Understanding how a threat model evolves over space and time is key to rescoping trust boundaries to cope. It is also critical to evaluate whether misalignment of administrative and trust boundaries leads to incorrect assumptions about the threats posed by insiders, whom we model here as those located within an administrative boundary but who ought not to be deemed trustworthy. Usability also matters: security controls can be too strict leading to circumvention or to a product being abandoned. Avoiding poor usability outcomes, and thus poor commercial outcomes means keeping trust boundaries and administrative boundaries realistically aligned. 

Future threat modelling work should consider how development teams can use tools and concepts such as $TM\textsubscript{$\Delta$}$ to map the gap between different types of boundary and maintain situational awareness of threats that arise because of features added elsewhere -- or indeed of changes in assumptions about the broader threat environment.

\bibliographystyle{IEEEtran}
\bibliography{e2ee}
\end{document}